\documentclass[12pt,preprint]{aastex}
\usepackage[onecolumn]{emulateapj5}
\usepackage{apjfonts}
\usepackage{psfig}
\shorttitle{Large Source Microlensing}
\shortauthors{Agol}

\begin{document}


\title{Microlensing of Large Sources}

\author{Eric Agol\altaffilmark{1,2}}
\altaffiltext{1}{Chandra Fellow}

\altaffiltext{2}{California Institute of Technology, Mail Code 130-33, 
Pasadena, CA 91125 USA; agol@tapir.caltech.edu}

\slugcomment{}

\begin{abstract}
We prove a gravitational lensing theorem:  the magnification of 
a source of uniform brightness by a foreground spherical lens 
is $\mu =1+\pi(2R_E^2-R_L^2)/A$, where $A$ is the area of the source 
and $R_E$ and $R_L$ are the Einstein radius and size of the lens 
projected into the source plane; this provides an accurate 
approximation to the exact magnification for $R_L^2,R_E^2 \ll A$.  
Remarkably, this result is 
independent of the shape of the source or position of the lens (except 
near the edges).   We show that this formula can be generalized to 
include limb-darkening of a circular source by simply inserting the 
surface-brightness at the position of the foreground object (divided 
by the average surface-brightness of the star).  We also show
that similar formulae apply for a point-mass lens contained in 
a shear field and mass sheet, and for an ensemble of point masses
as long as the Einstein radii are much smaller than the source size.
This theorem may be used to compute transit or microlensing lightcurves for which 
the foreground star or planet has a size and Einstein radius much
smaller than the background star.    
\end{abstract}

\section{Introduction}

With the advent of sensitive large-area synoptic surveys of the sky, it
may be possible in the near future to detect non-accreting compact objects 
in binaries via gravitational lensing during transit of the companion stars, either
main-sequence or white dwarf \citep{mae73,mar01,ago02,bes02,sah02}.  The problem
of modeling the lightcurve of a transiting star has been solved for
a uniform source by \citet{ago02}, while a limb-darkened source requires
numerical integration \citep{ago02,sah02}.  A related problem occurs in
the search for planetary companions of stars via perturbations to microlensing 
lightcurves which requires numerical solution for the planetary perturbation.  
Here we prove a theorem that
the solid angle added by a lens star is equal to $2\pi$ times the square of
the angular Einstein radius for lensing of a source which has an angular size 
much larger than angular Einstein radius of the lens ($\S 2$).  We 
analytically compute the transit lightcurve of a compact object in front
of uniform star, including eclipse by the lens ($\S 3$), which we use to derive
the lightcurve for a limb-darkened source star ($\S 4$), which we generalize 
to a lens with shear and convergence ($\S 5$) as well as the case of multiple lenses 
in front of a large source ($\S 6$), and finally we apply these formulae to several 
astrophysical problems ($\S 6$).  Capitalized $R$'s, such as the Einstein 
radius, $R_E=\left(4 G M c^{-2} (D_L^{-1}-D_S^{-1})\right)^{1/2}$, are in units of angular 
size, while lower-case $r$'s, such the source size $r_*$, are in units of $R_E$, 
$r_*=R_*/R_E$, where $D_L$ and $D_S$ are distances to the lens and source and
$M$ is the lens mass.

\section{Magnification of a large source}

We show first that the magnification of a large uniform source is approximately 
constant when the lens is in front of and away from the edges of the source.  
The lens equation normalized to $R_E$ is $y=x-1/x$ where $y$ is the source 
position and $x$ is the image position.  In the case of a lens in front of a 
uniform source of radius $R_*$ for which $R_E \ll R_*$ and for which the closest 
distance of the lens to the edge of the source is much smaller than the Einstein 
radius, then $y$ much larger than unity.  The magnification is the area contained within the 
outer lensed image divided the unlensed area, so $x\sim y+1/y$ is large 
as well  (the area within the inner image should in principle be subtracted
off, but it is of order $y^{-2}$ so that it can be ignored). 
Thus, the magnification is given by
\begin{equation}
\mu_e = {1\over A} \int_0^{2\pi} d\phi {1\over 2}x^2
\end{equation}
where $x$ is the distance to the images located at
a position angle $\phi$ relative to the lens center (see Figure 1) 
and $A$ is the area of the source in units of $R_E^2$. 
Squaring the lens equation gives $y^2=x^2-2+1/x^2$, but since $x$
is large, $x^2\simeq y^2+2$.  Thus
\begin{equation}
\mu_e = {1\over A} \int_0^{2\pi} d\phi {1\over 2}(y^2+2).
\end{equation}
Now, $\onehalf y^2 d\phi$ is simply the differential area of
the unlensed source, so the integral of this term gives $A$.
The remaining term is $d\phi$ which means that the additional
area due to lensing is constant per differential angle.  

Figure 1 shows two shaded regions between the source and image
boundaries subtended by the same angle
$\delta\phi$.  These shaded regions represent the extra solid angle
added by lensing.  Despite their different shape, the two
regions have the same area to order $y^{-2}$.  This
is simply due to the fact that more distant points $y$ have
smaller displacements $\Delta y\sim 1/y$ so that the total area, $dA=yd\phi \Delta y
\sim d\phi$ is simply equal to the subtended angle, independent
of the distance to the boundary or shape of the boundary.
Thus, the total of the extra area is simply $2\pi$ (in units 
of $R_E^2$) independent of the size or shape of the source or
the position of the lens (as long as it is away from the
edges).  The maximum area added is equal to $2\pi$ since for
sources small compared to an Einstein radius, $r_* \ll 1$,  the 
extra area is $2\pi r_* \ll 2\pi$ (however, the magnification is
larger for small sources since the unlensed flux is proportional
to $r_*^2$).  So the total magnification for a large source then becomes
\begin{equation}
\mu_e = 1 + {2\pi \over A}.
\end{equation}
\citet{mae73} has derived a similar result for the special case of a
circular source.
So we have proved that the magnification of a lens in front of a large 
source is constant to order $y_{min}^{-2}$,
where $y_{min}$ is the shortest distance of the center of the lens
to the edge of the source in units of $R_E$.  This creates a lightcurve
which looks like a top hat.  

For a circular source,
$A=\pi r_*^2 = \pi (R_*/R_E)^2$, so $\mu_e= 1 + 2(R_E/R_*)^2$.
When the lens is outside of the source, one can show that
there is no extra magnification (the area is the difference of
the inner and outer images of the source, so the factor of 2 cancels
when taking the difference), so to a good approximation the magnification 
is simply 
\begin{equation}\label{step}
\mu_e = 1+\left[2(R_E/R_*)^2-(R_L/R_*)^2\right]\Theta(r_*-z),
\end{equation}
where $z$ is the position of the lens with respect to the origin
of the circular source and $R_L$ is the angular size of the lens
star.

\centerline{\psfig{file=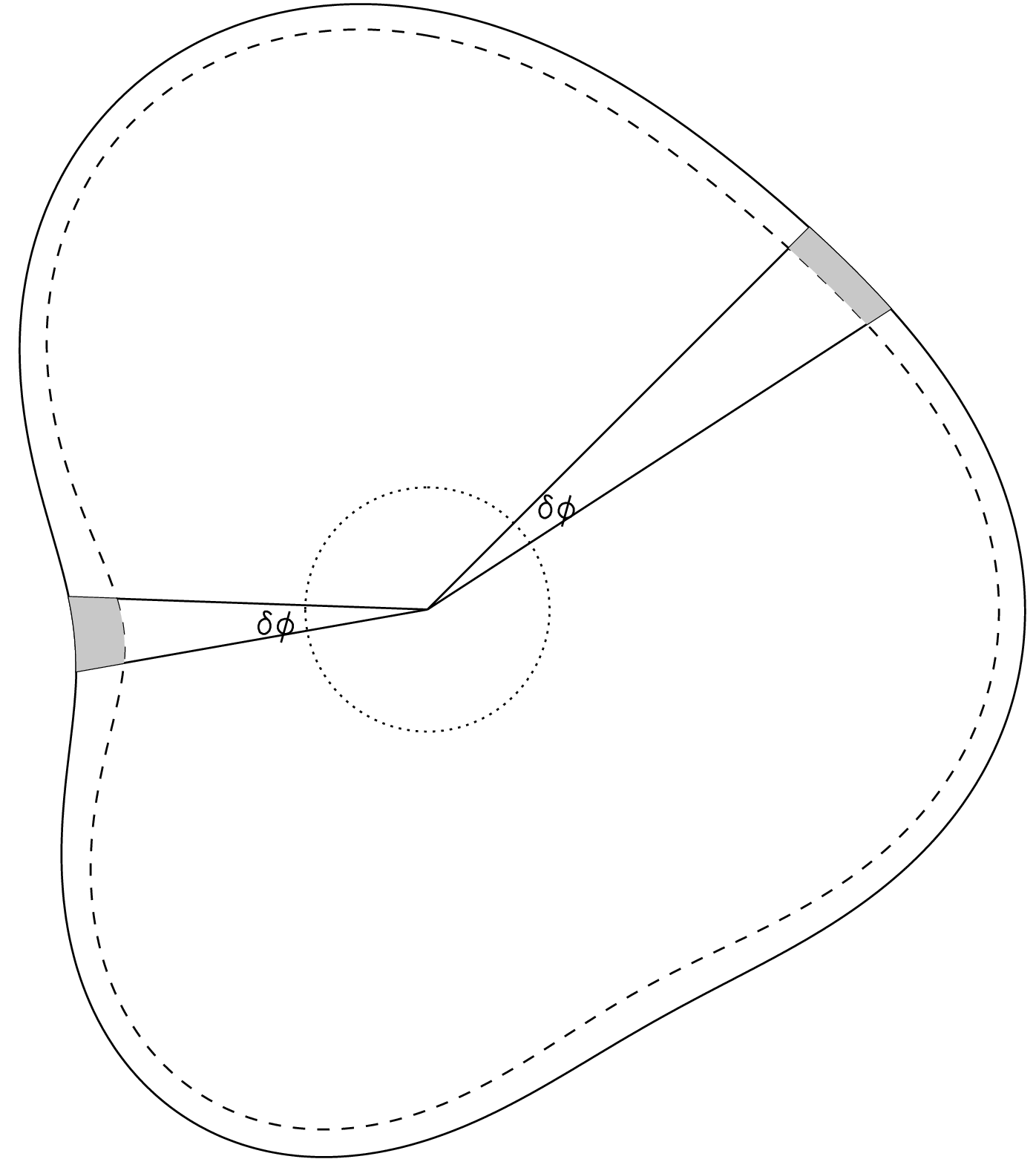,width=3in}} 
\figcaption{Magnification of a large source by a lens with
small Einstein radius.  The dotted line is a circle with radius
$R_E$ (or unity in normalized units).  The dashed line is the
boundary of the unlensed source, while the solid line is the
boundary of the lensed source.}

\section{Magnification of a large source near the edge}

To compute the magnification of an extended source near the
edge, we cannot ignore the $x^{-2}$ term in the squared lensing
equation, so we must include the distortion of the edge of the 
star.  In the case of a large circular star, $R_* \gg R_E$, the edge 
can be approximated as a straight line and the change in magnification
is simply the extra area enclosed by the image of this line.
Figure 2 shows the geometry of a line being lensed by a point-mass
lens.   As the lens approaches the line, it is distorted outward,
curving around the Einstein ring.  The inner image of the line
grows inside the Einstein ring, filling half of the Einstein ring 
when the line crosses the origin.  Then, as the line crosses to the 
other side, the pattern repeats itself as the line recedes from the
center of the disk.  This creates a symmetry about the point
at which the lens sits on the limb of a large 
source.  The magnification is given by
\begin{equation} \label{edge}
\delta \mu_e = {1\over r_*^2} -{2(z-r_*) \over \pi r_*^2 k}
\left[K(k)-E(k)\right],
\end{equation}
where $k^{-1}=\sqrt{1+(z-r_*)^2/4}$.  This approximates the
exact magnification quite well in the large-source limit.
Unfortunately, this expression is complicated enough that
we cannot integrate it over limb-darkening analytically.

\centerline{\psfig{file=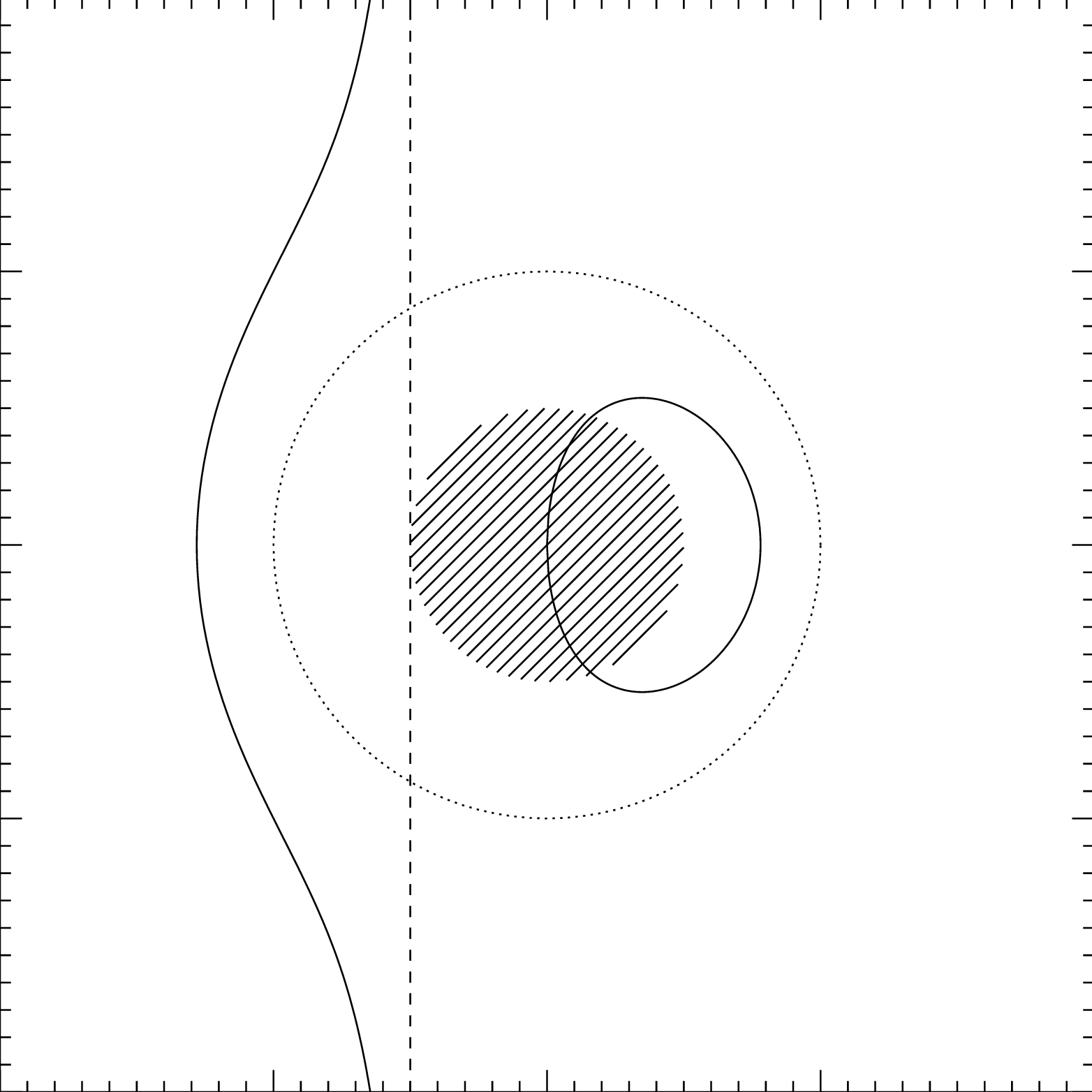,width=3in}} 
\figcaption{Magnification of a straight line by a point mass lens.
The dashed line is the unlensed line, the dotted line is
the Einstein ring, the solid lines are the images of the
line.  Note that the images are the same as the paths of
a point source moving in a straight line behind a lens. The shaded
region represents the area blocked by the lens of finite
extent $r_L<1$.}

Figure 3 compares equation (\ref{edge}) with the exact equations
from Agol (2002).  For smaller source size, the edge has a
Napoleonic bicorne shape, while larger sources look more like
a tophat.  The straight-edge approximation improves as the source
becomes larger, and approaches a step function as $R_*/R_E 
\rightarrow \infty$.

\centerline{\psfig{file=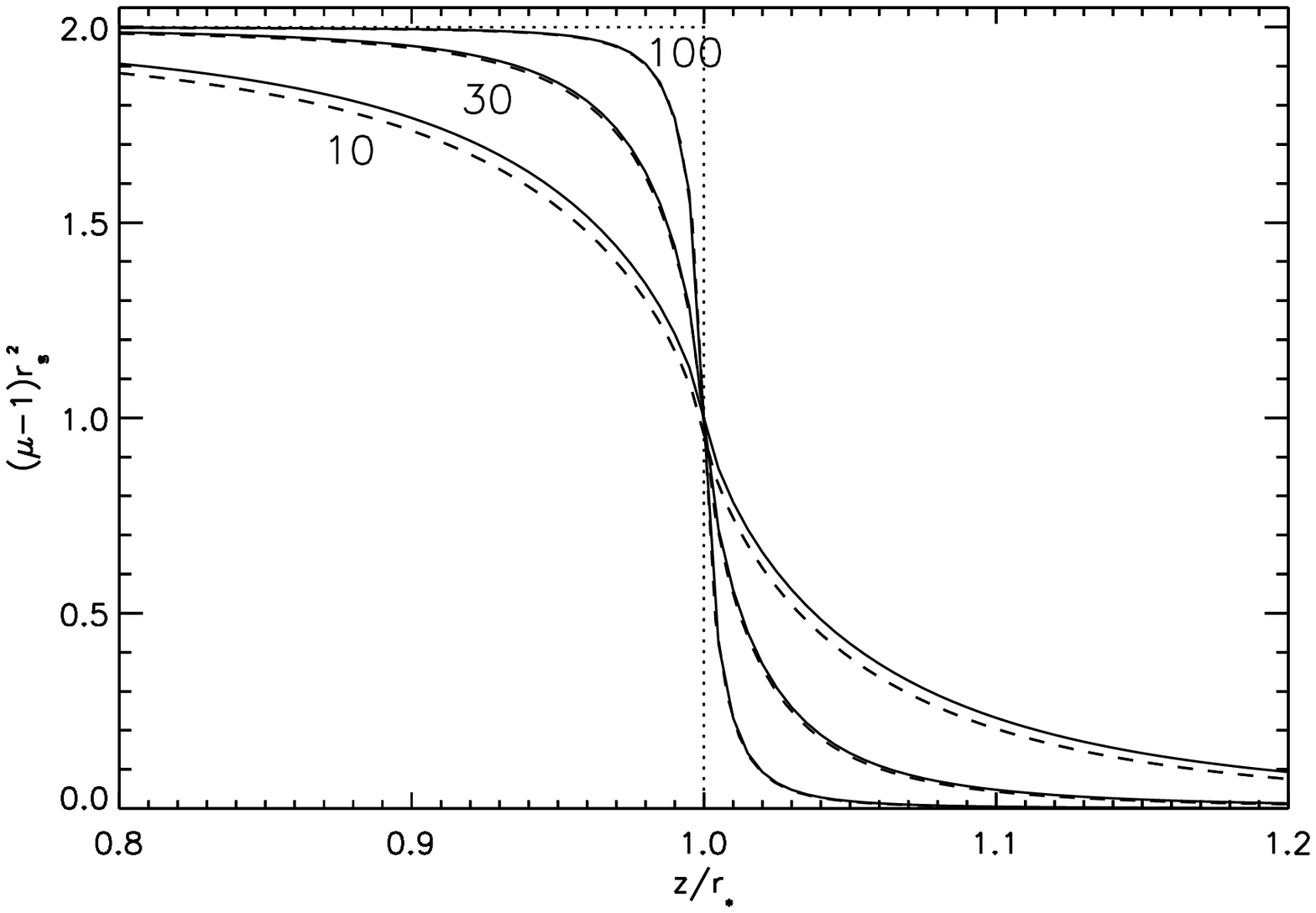,width=5in}} 
\figcaption{Magnification of a large, circular uniform source 
by a lens near the edge as described by equation (\ref{edge}) 
(solid lines) or the exact equation (dashed lines).  The 
horizontal axis is the separation of the lens and source in
units of the source radius.  The curves are labeled by the
value of $R_*/R_E$, while the dotted line is equation (\ref{step}).
The magnification plotted is
subtracted from unity and multiplied by $(R_*/R_E)^2$ for better
comparison between the various source sizes.}

When the lens has a finite size, $r_L=R_L/R_E > 0$, the obscuration of the
images by the lens gives
\begin{eqnarray}\label{microccult}
\delta \mu &=& {1\over r_*^2} +{B \over \pi r_*^2},\cr
B&=& s_\zeta \left({\pi \over 2}-\phi_0\right)(1-r_L^2)-{\pi r_L^2 \over 2}
+G\left({\pi\over 2}\right)-s_rG\left(\phi_0\right)
-\zeta \onehalf \tan{\phi_0} \left[\vert\zeta\vert+s_r \sqrt{\zeta^2
+4\cos^2{\phi_0}}\right],\cr
G(\phi)&=& {\zeta \over k}\left[K(\phi,k)-E(\phi,k)\right],\cr
\phi_0 &=& \left\{\begin{array}{ll} 0 & \vert\zeta\vert r_L \ge \vert1-r_L^2\vert \cr 
\cos^{-1} \left({\vert\zeta\vert r_L \over \vert 1-r_L^2\vert}\right) & 
\vert\zeta\vert r_L < \vert 1-r_L^2\vert,\end{array}\right.
\end{eqnarray}
where $\zeta=r_*-z$, $s_\zeta = sign(\zeta)$, $s_r=sign(r_L-1)$.

If an observed microlensing (or occultation) event of a uniform
source has a maximum depth given by $\delta \mu_{max}$, and if the 
limb-darkening function is known, then the measured depth can
be fit by any $R_L=\left[2R_E^2-\delta\mu_{max}R_*^2\right]^{1/2}$.  However,
the lightcurve near the edge of the source depends differently
on $R_E$ and $R_L$.  In Figure 4, we show lightcurves near the
edge computed from equation \ref{microccult} which have the same 
$\delta\mu_{max}$, but have different
ratios of $R_E/R_*$ and $R_L/R_*$.  The cases with larger Einstein
radii have a larger magnification and dip near the edge than the
pure occultation case.   The peak in $\delta\mu$ corresponds
to the point at which the image outside of the Einstein radius
is first occulted, at which point $z-r_*=r_L-1/r_L$ for $r_L>1$.
At this point, the magnification is
\begin{equation}
\delta\mu_{peak} = {1 \over 2 r_*^2}\left[1 - {r_L^2-r_L^{-2} \over \pi }\left\{K\left({2 \over r_L+r_L^{-1}}\right)
-E\left({2 \over r_L+r_L^{-1}}\right)\right\}\right].
\end{equation}
In the limit that $r_* \gg r_L \gg 1$, this reduces to
$\delta \mu_{peak}=4^{-1}r_*^{-2}r_L^{-2}$, which becomes small
rapidly.  In principle this can be used to break the degeneracy 
between $R_E$ and $R_L$ since the maximum magnfication has a 
different dependence on these quantitities.

\section{Lensing of a large limb-darkened source}

When limb-darkening is included using
the parameterization $I(R)=1-\gamma_1(1-\mu)-\gamma_2(1-\mu)^2$
and $\mu=[1-R^2]^{1/2}$, the
magnification outside of eclipse is decreased slightly due to
the darker limb, while the dip within the eclipse disappears
since the dip becomes larger as the lens covers the brighter regions
of the star (Figure 4).  These deviations may be useful for distinguishing a 
brown dwarf from a planetary transit, or a close white dwarf
from a terrestrial planet.

\centerline{\psfig{file=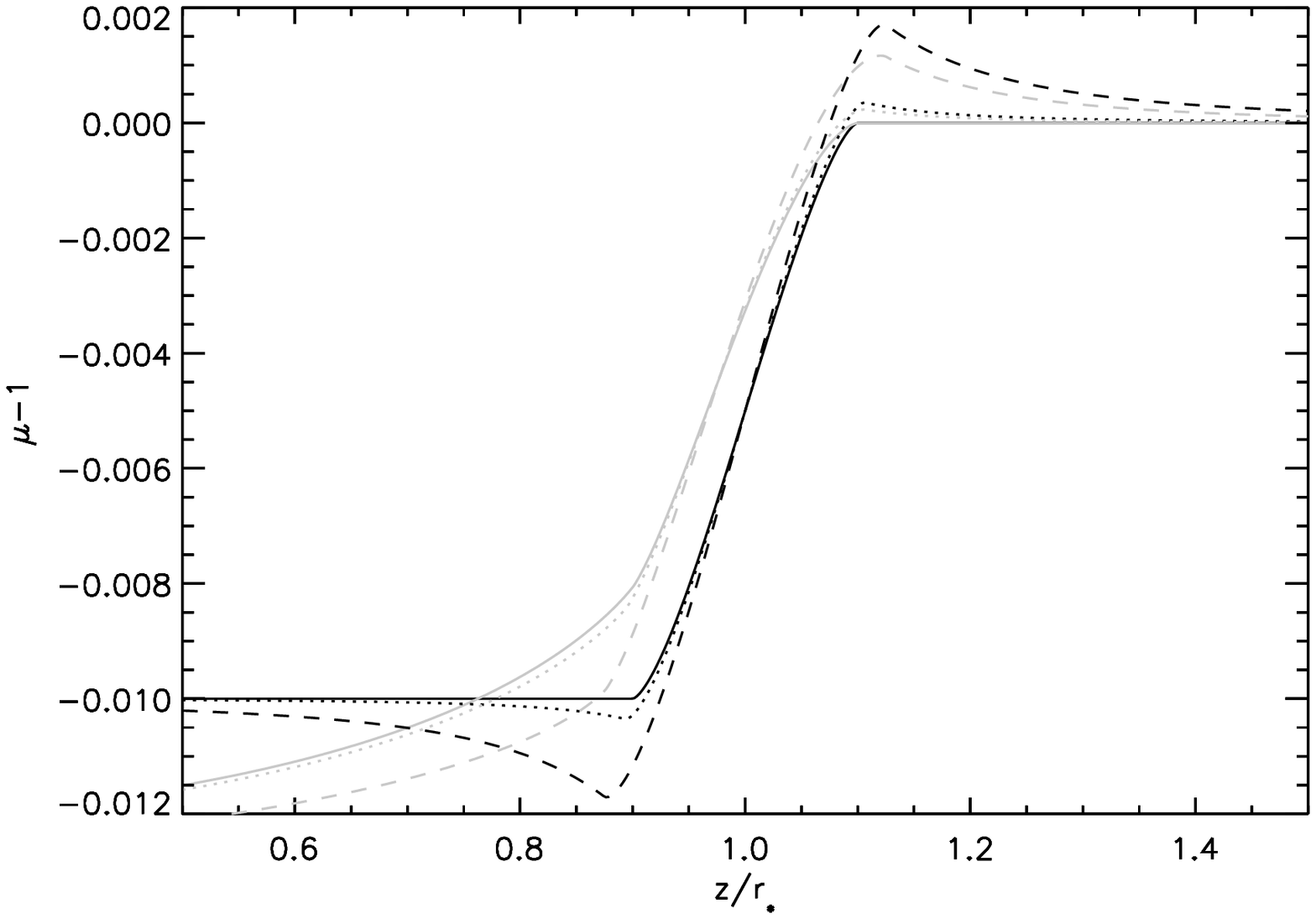,width=5in}} 
\figcaption{Magnification near the edge for $\delta\mu_{max}=-0.01$
for $R_E=10^{-2}R_*$, $R_L=0.1R_*$ (solid curve), $R_E=0.00707R_*$, 
$R_L=0.01414R_*$ (dotted curve), and $R_E=0.122474R_*$, $R_L=0.2R_*$ (dashed curve).
The lighter curves show the same parameters with quadratic
limb-darkening of $\gamma_1=0.3$ and $\gamma_2=0.3$ (Agol 2002).}

In the case of a limb-darkened star, we must integrate the magnification
for a uniform source over the limb-darkened profile of the star.
The magnification of a limb-darkened source is approximately
\begin{equation}
\mu_l = 1+ F_0^{-1} \int_0^{R_*}2\pi dR 
{d[\delta \mu_e(R) R^2]\over dR}I(R),
\end{equation}
where $F_0=2\pi\int_0^{R_*} dR R I(R)$ is the total flux of the source
star and $\delta \mu_e(R) = \mu_e(R)-1 
=2(R_E/R)^2\Theta(R/R_E-z)$.  Since $\delta \mu_e R^2$ is simply
proportional to the step function, its derivative is the delta
function.  Integrating over this delta function picks out the
surface brightness at the position of the lens.  We finally find that
\begin{equation} \label{approx}
\mu_l = 1+  \left[2\left({R_E\over R_*}\right)^2-\left({R_L \over R_*}\right)^2\right] 
{I(zR_E)  \over \langle I \rangle}\Theta(R_*-zR_E),
\end{equation}
where $\langle I \rangle = F_0/(\pi R_*^2)$ and we have added in
occultation by the foreground star.
It is remarkable that although the magnification involves integrating
over the surface of the entire star, it ends up depending
only on the surface brightness underneath the lens!  This results from
the fact that the microlensing amplification of a point source becomes
infinite for an exactly aligned source, which may be approximated as 
a delta function.  

Figure 5 shows a comparison of this approximation with exact lightcurves.
The agreement between the approximate lightcurves and exact lightcurves
away from the edges is astoundingly good, especially considering that the 
analytic approximation involves a great deal less computational time.
This formula gives a very good approximation to the 
magnification of a source with arbitrary limb-darkening for $R_E < 0.1 R_*$.  

\centerline{\psfig{file=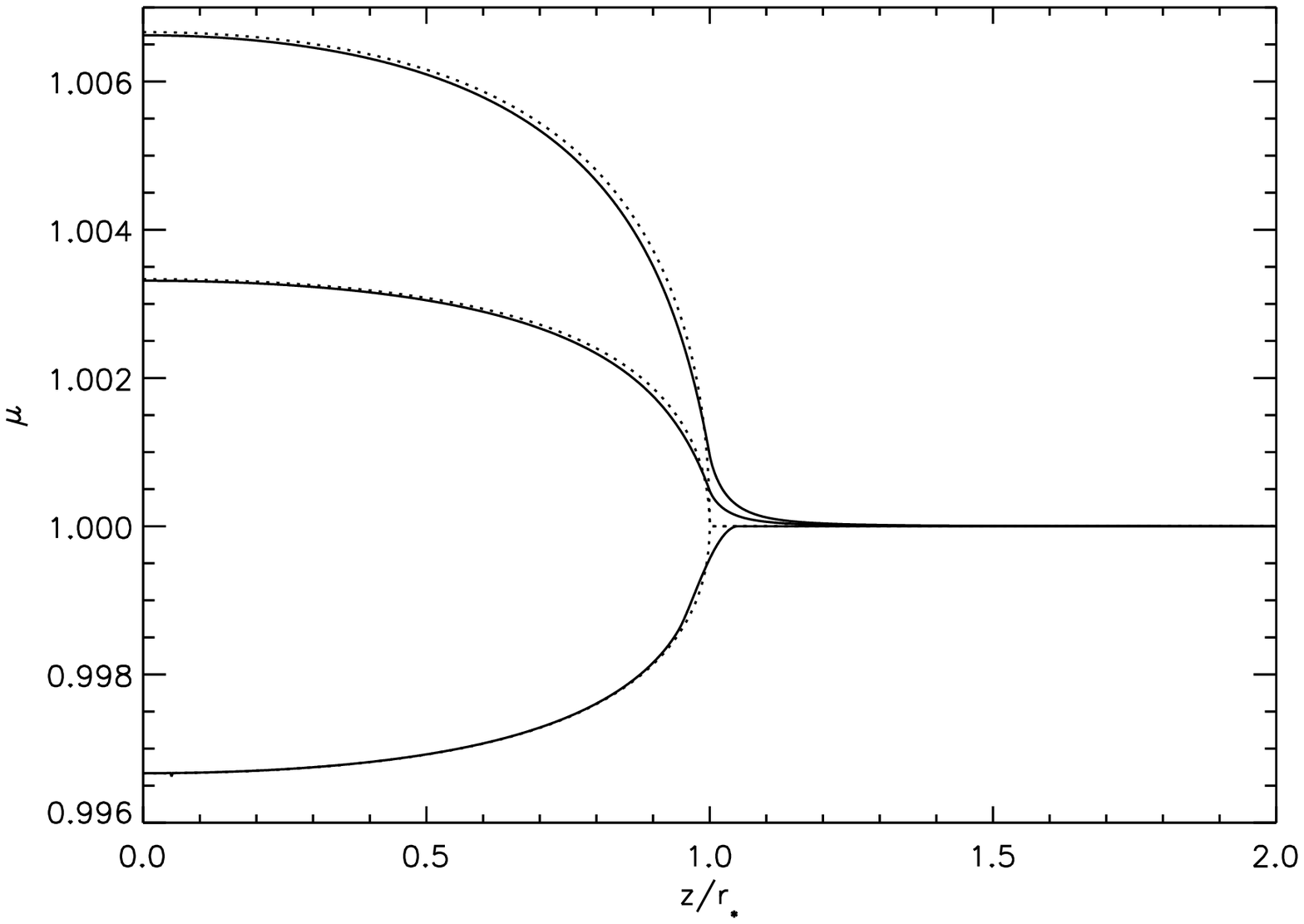,width=5in}} 
\figcaption{Magnification of a large source by a lens with
$R_*=20 R_E$ and $R_L=0$ (top curve), $R_*=20 R_E$ and
$R_L=R_E$ (middle curve), and $R_*=20 R_L$ and $R_E=0$
(bottom curve) from \citet{ago02} for $\gamma_1=\gamma_2=0.3$.  The 
dotted lines are the approximation given by equation (\ref{approx}).}

\section{Magnification of a large source by a Chang-Refsdal lens}

We now generalize the theorem to the case of a point-mass lens
embedded in a field with constant shear and convergence.  A 
shear field distorts circles into ellipses, while a convergence
field simply expands the size of a source,  and since lensing by a 
point mass creates extra area equal to $2\pi$ independent of source size
or shape, one might expect that
the constant magnification should still apply.  The Chang-Refsdal lensing 
equation is
\begin{equation} \label{crlens}
{\bf y} = {\bf x} \left(1-{1\over x^2}-\kappa\right) +\gamma \left(
\begin{array}{cc}
1 & 0 \cr
0 & -1
\end{array}
\right)
{\bf x},
\end{equation}
where ${\bf x} =(x_1,x_2)$ and ${\bf y}=(y_1,y_2)$ are the
image and source coordinates, respectively, $\kappa$ is the convergence 
or normalized surface density,
and $\gamma$ is the shear \citep{sch92}.   The coordinates have been chosen
such that the shear axes lie along the coordinate axes.
Now, the area of the image is given by the closed integral
$A_x = \onehalf \int {\bf x}\times d{\bf l_x}$, where $d{\bf l_x}$
is the line element for the contour described by ${\bf x}$, and 
similarly for the source replacing $x$ with $y$.
One can use the lens equation (\ref{crlens}) to find the zeroth
order solution for ${\bf x}$ which amounts to neglecting bending
by the lens
\begin{eqnarray}
x_{01} &=& y_1 \left(1-\kappa+\gamma\right)^{-1},\cr
x_{02} &=& y_2 \left(1-\kappa-\gamma\right)^{-1}.\cr
\end{eqnarray}
Then, the bending angle can be approximated as ${\bf x}/(x_{01}^2+x_{02}^2)$,
giving a first-order solution
\begin{eqnarray}
x_1=y_1\left(1-{1\over x_0^2}-\kappa+\gamma\right)^{-1},\cr
x_2=y_2\left(1-{1\over x_0^2}-\kappa-\gamma\right)^{-1},
\end{eqnarray}
where $x_0^2=x_{01}^2+x_{02}^2$.  Using this solution, one can compute the
area of the lensed image, $A_x$.  For a lens at the origin, the magnification is
\begin{equation}
\mu_{CR}(0) = \mu_0 + {2\Theta(\vert 1-\kappa\vert-\gamma) \over (1-\kappa) r_*^2},
\end{equation}
where $\mu_0=\vert(1-\kappa)^2-\gamma^2\vert^{-1}$.
This is a generalization of the formula in \citet{gou97} which
was derived for $\kappa=0$.  For a lens located off of the origin, it can
be shown numerically that the magnification is approximately constant for
a lens within a large source, and then approaches $\mu_0$ outside of the
source, so the extra magnification can be multiplied by a step function as in
the point-mass case, and the specific intensity can be included as well.
This gives
\begin{equation} \label{cramp}
\mu_{CR}(z) = \mu_0 + {I(zR_E)\over \langle I\rangle}{2\Theta(\vert 1-\kappa\vert-\gamma) \over (1-\kappa) r_*^2}
\Theta(r_*-z).
\end{equation}
Figure 6 shows a comparison of lightcurves for a Chang-Refsdal lens passing
in front of a source of size $r_*=10$ for different values of $\kappa$ and $\gamma$.
Note that if $\vert 1-\kappa\vert < \gamma$, then the lightcurve is unperturbed.  The
magnification differs strongly near the edges due to the presence of caustics, 
but otherwise does an excellent job of describing the microlensing.

\centerline{\psfig{file=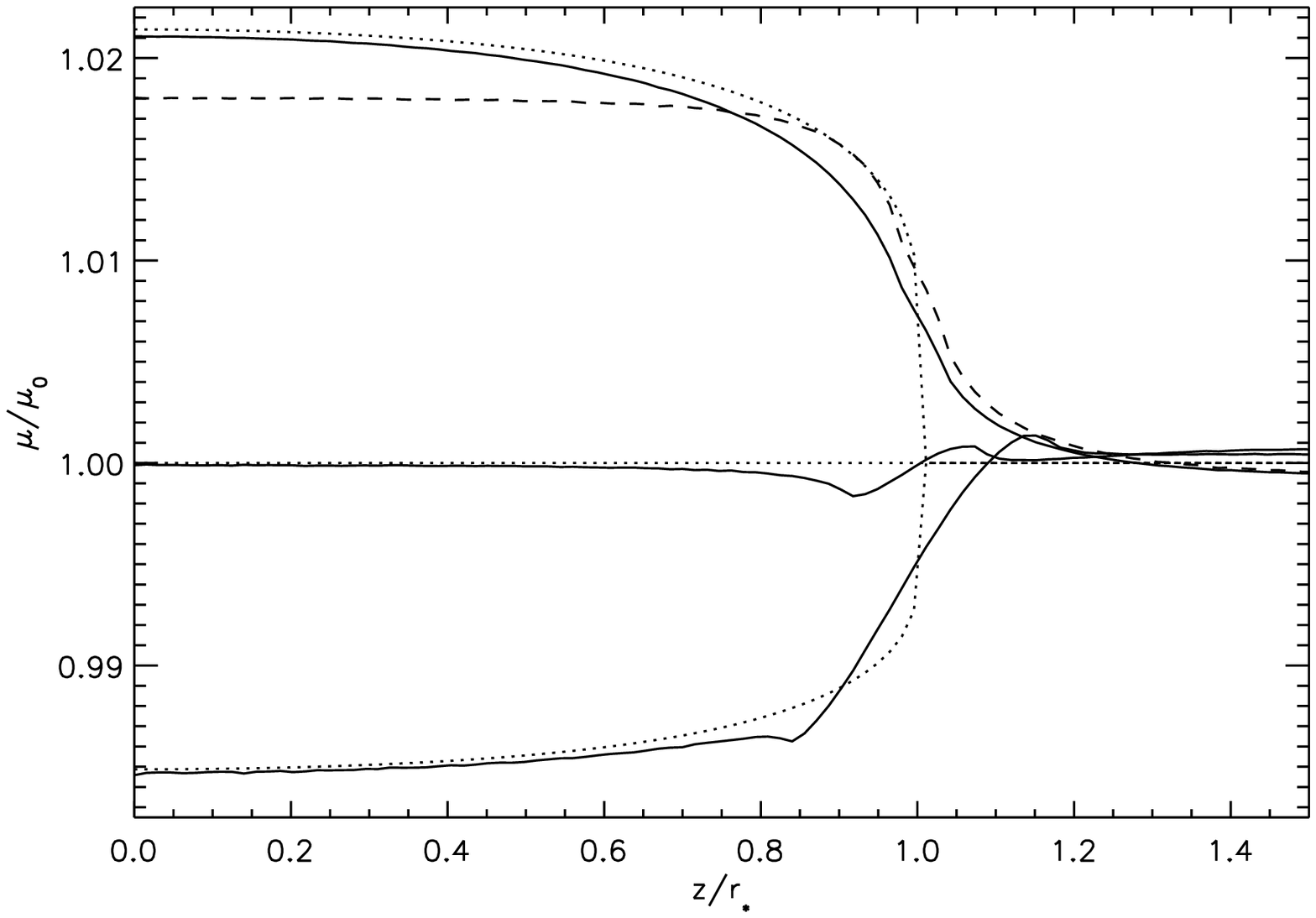,width=5in}} 
\figcaption{Magnification of a limb-darkened source with limb-darkening coefficients
of $\gamma_1=0.3$, $\gamma_2=0.3$.  The top solid curve is for $\kappa=0, \gamma=0.3$,
the middle curve is for $\kappa=0.9, \gamma=0.$ (since $\vert 1-\kappa\vert <\gamma$, there
is no perturbation), and the bottom curve is for $\kappa=1.7, \gamma=0.2$.  The
dashed curve is the same as the top curve without limb-darkening.  The dotted
lines show the analytic approximation given by equation (\ref{cramp}).  In
each case the lens starts at the center of the source and moves at a 45$^\circ$
angle with respect to the lensing axes as determined by the direction of the shear.}

\section{Magnification of a large source by many lenses}

We next generalize the theorem to the case of lensing by a 
distribution of point masses.  In this case the normalized
lens equation is
\begin{equation}
{\bf y} = {\bf x} - \sum_i m_i {{\bf x} -{\bf x_i} \over \vert {\bf x} -{\bf x_i}\vert^2},
\end{equation}
where $m_i$ is the fraction of the total mass in each lens ($\sum_i m_i=1$)
and ${\bf x_i}$ is the position of each lens.  As in the single-lens
case, we can square this equation, disregard terms of order $x^{-2}$, and
rearrange terms to give
\begin{equation}
y^2 = x^2 - 2 - 2\sum_i m_i {{\bf x_i}\cdot ({\bf x}-{\bf x_i}) \over  \vert{\bf x} -{\bf x_i}\vert^2}.
\end{equation}
This looks similar to the single-lens case except for the extra term which
appears since each lens is not at the center of the coordinate system.  One
can approximate ${\bf x} \sim {\bf y}$ and then carry out the integral over
the last term.  In the case of a circular source, this term gives zero.
So, we end up deriving the same result as in the single-lens case, equation
(\ref{step}),  the only difference is that the Einstein radius is determined
by the sum of all masses in front of the lens; the condition $R_E \ll R_*$
must still hold.  

Now consider a random star field with a surface number density of stars
$n_*$.  The lensing optical depth is defined as $\tau_* = \pi R_{E*}^2 
n_*$, where $R_{E*}$ is the Einstein radius for each star.  The mean
number of stars in front of a large source is then $N_*=n_* \pi R_*^2$, so
if $R_E=\sqrt{N_*} R_{E*} \ll R_*$, or $\tau_* \ll 1$, then we can apply 
the above result
to find the mean magnification:  $\langle \mu\rangle = 1 + 2 \tau_*$.
Of course, a random distribution of stars will create Poisson fluctuations
in the number of stars in front of the source.  Thus, the standard
deviation of the magnification for an ensemble of star fields is
$\sigma_\mu = 2 \sqrt{\tau_*} R_{E*}/R_*$.  This result has already been derived
by \citet{ref91} using a different computational technique.

\section{Applications}

The formulae derived here will be useful for discovering compact objects
in eclipsing binaries.  A measurement of lensing during a transit by a
compact object of a companion star places a constraint on the lens
mass \citep{sah02,bes02}.  Brown dwarfs (BD), white dwarfs (WD), black 
holes (BH), and neutron stars (NS) in binaries of semi-major axis $a$ each 
have Einstein radii, $R_E=\sqrt{4GMc^{-2}a}$,
much smaller than the size of a companion main-sequence star (MS) for
$a \sim 1$ AU and $M\sim 1M_\odot$,
and thus the approximation of equation (\ref{approx}) will apply.
Figure 8 shows an example lightcurve of a white dwarf orbiting a main
sequence star for $M_{WD}=0.7 M_\odot$, $M_{MS}=1 M_\odot$, $a=0.1$ AU, 
$t_{WD}=10$ Gyr, $\gamma_1=0.6, \gamma_2=0$, and the observer in
the equatorial plane of the binary.  In the upper left panel, the analytic 
approximation (equation \ref{approx}) is plotted as a dotted line, which 
is barely distinguishable from the exact lightcurve (solid line).  The transit/eclipse 
duration is about 3 hours.  There is slight tidal
distortion of the main-sequence star with an amplitude of 
$\sim 5\times 10^{-5}$ and sinusoidal dependence of twice the frequency
of the binary.  In this particular example,
$r_L=1$ and $r_S=88$, or the Einstein radius is $R_E=0.011
R_\odot$.  The V-band magnitude of the WD is $M_{V,WD}=17$, while
the MS star is $M_{V,MS}=4.8$, which makes the eclipse of the
white dwarf by the main-sequence star (at 160 hours) rather faint.
The maximum deviation of the lightcurve is to be $\sim 8\times 10^{-5}$,
this is modified somewhat due to limb-darkening.
To detect such a transit requires better than $0.01$ millimagnitude sensitivity
over several hours, which is a requirement that should be met by the
planned {\em Kepler} and {\em Eddington} missions.

\centerline{\psfig{file=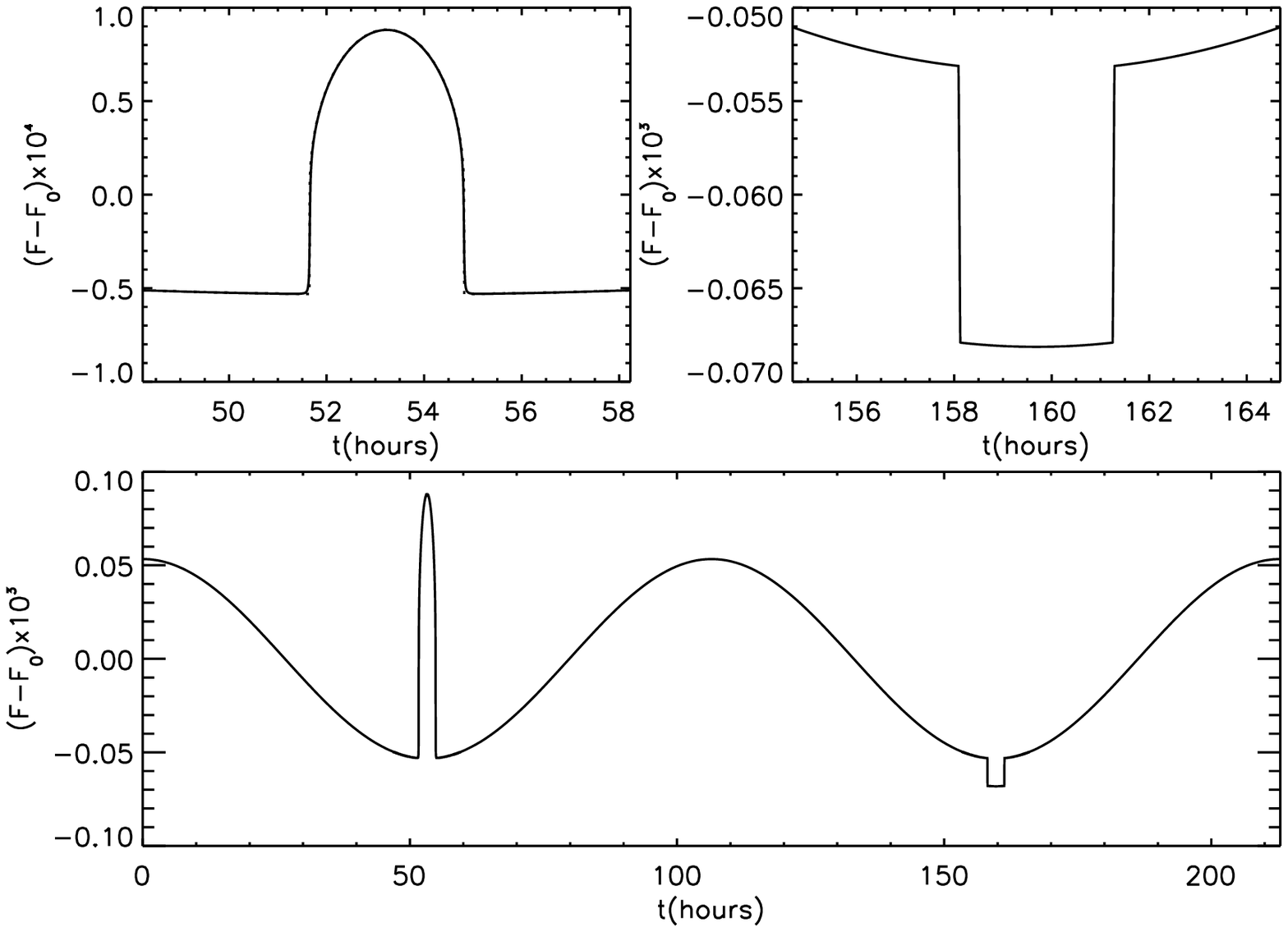,width=5in}} 
\figcaption{V-band lightcurve of a WD-MS binary described in
the text.  The
upper panels show the region near the white dwarf transit
(left) and the white dwarf eclipse (right).  The full
lightcurve (213 hours) is shown in the bottom panel.}

A simple estimate of the size of the lensing signal can be made
in the limit that the size of the lensing object can be neglected (usually
the case for neutron stars, black holes, and white dwarfs with large semi-major
axes).  The extra flux caused by lensing
is simply $2 r_*^{-2} F$ (neglecting limb-darkening), where $F$ is the flux of the unlensed
companion star.  The duration of transits is of order $2 R_E/v$, where
$v$ is the relative velocity of the binary, while the number of transits
scales as $T_{obs}/P=T_{obs}v/(2\pi a)$, where $T_{obs}$ is the total
observing time (assumed to be much larger than the period of the binary
and the monitoring is assumed to be continuous),
$a$ is the semi-major axis of the binary (assumed to be circular). Then
the total fluence added by the lensing is equal to the lensing flux times the
transit duration times the number of transits, or
\begin{equation}
\delta E = 2 F {R_E^2 \over R_*^2} {2 R_E \over v} {T_{obs} v \over 2\pi a}
= {8\over \pi} {R_g \over R_*} F T_{obs},
\end{equation}
where $R_g = GM_{CO}/c^2$ is the gravitational radius of the lensing compact
object.  Remarkably, any dependence on the semi-major axis or mass of
the main-sequence star has disappeared!
This is due to the fact that more distant binaries have a larger flux signal 
and longer transit duration, but fewer transits during a given experiment duration.
Since $E=F T_{obs}$ is the fluence of the unlensed star, the total
signal is 
\begin{equation}
{\delta E \over E} = {8\over \pi}{R_g\over R_*} = 5 \times 10^{-6} {m \over R_{*\odot}},
\end{equation}
where $m$ is the mass of the lensing, compact object in solar masses and
$R_{*\odot}$ is the radius of the main-sequence star in units of solar radii.
This relation suggests a technique for measuring the mass of the eclipsing object:
first, estimate the radius of the main-sequence companion via spectral or photometric
modeling (the distance can be unknown), and then measure $\delta E/E$ to derive 
$m=2\times 10^5 R_{*,\odot} \delta E/E$.

The foregoing formula is only true for an equatorial transit of a main-sequence
star in a band which is not limb-darkened.  Including limb-darkening and a non-equatorial
transit, one finds
\begin{equation} \label{fluence}
{\delta E \over E} = {8\over \pi}{R_g\over R_*}{2x\langle I \rangle_{line} \over \langle I \rangle_{disk}},
\end{equation}
where $x=\sqrt{1-b^2}$, $b$ is the minimum impact parameter of the lens with
respect to the center of the source star in units of $R_*$, $\langle
I \rangle_{line}$ is the mean surface brightness of the source along the path
of the lens, and $\langle I \rangle_{disk}$ is the mean surface brightness
of the source.  For quadratic limb-darkening, we find
\begin{equation} \label{blimb}
H(b)={2x\langle I \rangle_{line} \over \langle I \rangle_{disk}}
= {(1-\gamma_1-\gamma_2)x + {\pi \over 4} (\gamma_1+2\gamma_2)x^2-
{2\over 3} \gamma_2 x^3 \over \pi \left(1-{1\over 3} \gamma_1
-{1\over 6} \gamma_2\right)}.
\end{equation}
If $R_*$ and $\gamma_1,\gamma_2$ are known from, say, modeling of the spectrum of 
the main-sequence star, then equation \ref{fluence} has two unknowns,
$b$ (or $x$) and $M_{CO}$.  Another equation can be obtained from the
ratio of the duration of the transit, $t_{transit}$, to the period of the
orbit, $P$.  Since $P$ depends on the semi-major axis, which depends on the
masses of both stars, one finds
\begin{equation}\label{timescales}
\sqrt{1-b^2} R_* = t_{transit} \left[2\pi G P^{-1}\left(M_{MS}+M_{CO}\right)\right]^{1/3},
\end{equation}
which constitutes another equation for $b$ and $M_{CO}$ assuming that $M_{MS}$
is known.  The mass of the compact object may be found by eliminating $M_{CO}$
in equations \ref{fluence} and \ref{timescales}, which yields a 6th order
equation for $x$ which can be solved numerically for $x$ which may then
be used in equation \ref{fluence} to find $M_{CO}$.   If the data are of
sufficient quality then $b$ may be found by fitting the shape of the transit
and then $M_{CO}$ may be found directly from equation \ref{fluence} (this
is described in \cite{man02} for the occultation case).
Uncertainties in $M_*, R_*, \gamma_1,$ and $\gamma_2$ will contribute to 
uncertainty in $M_{CO}$ and $b$.  Another technique for finding $b$ relies
on measuring the difference of the transit depth in different bands.
Figure 7 shows the ratio of $H(b)$ for the V and I bands as a function of
impact parameter for a solar-type star.  The variation is nearly a factor
of 3 for $b$ ranging between 0 and 1.  Given this ratio, $x$ may be
solved for using the quadratic equation in $x$ determined by 
equating the functions \ref{blimb} for V and I with the observed value.
The value of $b$ measured from
this relation can then be used to estimate $H(b)$ and determine $R_g/R_*$.

\centerline{\psfig{file=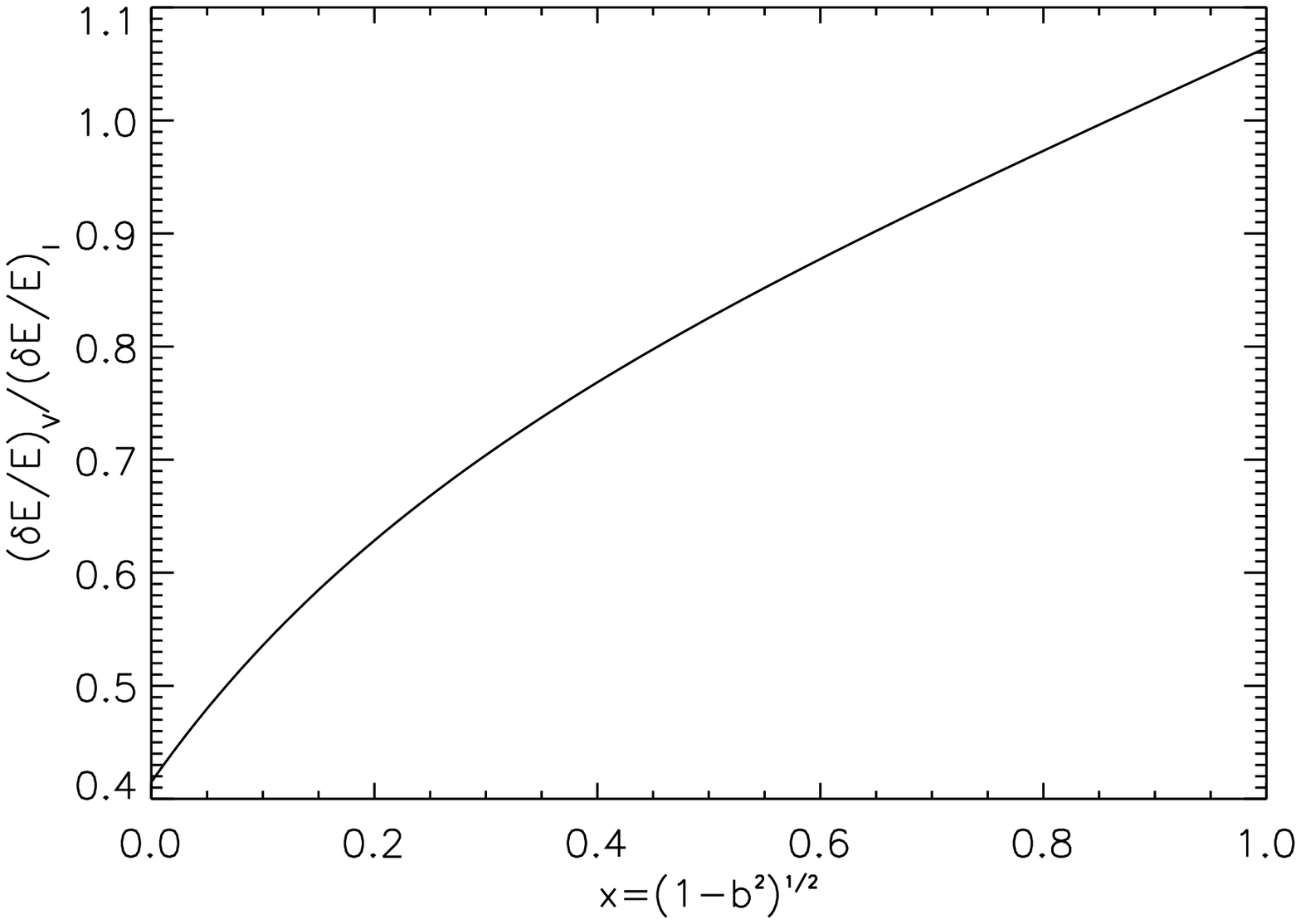,width=5in}} 
\figcaption{The function $H_V(b)/H_I(b)$ plotted versus $x=\sqrt{1-b^2}$
for a main-sequence G star.}

The transit detection probability is $R_*/a$, from which we can estimate
the fraction of binaries for which transits can be detected assuming the
the required fluence sensitivity can be achieved.   This is considered
in detail for WD-MS binaries by \citet{far03} who find that $>10^2$ such
systems might be found with the {\em Kepler} and {\em Eddington}
missions, and for WD-WD, WD-NS, WD-BH binaries by \citet{bes02}.   
Here we make some general considerations
for the detection of NS-MS and BH-MS binaries.  For the binary to not accrete, the
main-sequence star must have a size smaller than the Roche-lobe size, $R_L$,
by some factor $f_R=R_*/R_L$.  To a good approximation, the Roche-lobe is
given by $R_L=0.5 a f_M^{1/3}$ where $f_M=M_{MS}(M_{MS}+M_{CO})^{-1}$ and
the $CO$ subscript stands for compact object \citep{pri85}.  Typically, 
binaries show a distribution of semi-major axes which is constant in $\ln{a}$, 
which we assume to be the case from $a_{min}$ to $a_{max}$ for BH-MS and NS-MS
binaries (clearly, reality is likely to be more complicated due to the effects of
kick-velocities during the formation of the compact object, e.g. in a supernova
collapse), common-envelope evolution of the massive star, and tidal evolution
due to distortion of the main-sequence star).  Since $a_{min}$ is set by
the Roche-lobe overflow limit, $a_{min} = 2 R_* f_R^{-1} f_M^{-1/3}$,
the probability for lensing scales as
\begin{equation}
p_{lens}=R_*/a_{min} = 0.5 f_R f_M^{1/3} [ln(a_{max}/a_{min})]^{-1}.
\end{equation}
Note that since $a_{min} \propto R_*$, the probability is independent
of the size of the star!  For a population of solar-mass stars orbiting
10 $M_\odot$ black holes with, say, $a_{max}=10^5 a_{min}$ and $f_R=0.5$,
the probability of lensing is about 1\%.  The detected stars will be
heavily concentrated toward $a_{min}=9 R_{\odot}$ for this choice
of parameters, so $\delta \mu_e \sim 10^{-3}$.

The total number of non-accreting BH-MS or NS-MS binaries in our galaxy is quite
uncertain.  \citet{tut02} estimate that $(3-6)\times 10^4$ BH-MS
binaries and $(1-3)\times 10^5$ NS-MS binaries reside in the Milky way,
while \citet{lip96} estimate $\sim 10^{2.8}$ BH-MS and $\sim 10^4$ 
NS-MS binaries.  Given the above assumptions for $f_R$ and $f_M$, there exists
a total of $\sim 10^{2-3.5}$ NS-MS transiting
binaries and $\sim 10^{0.6-2.6}$ BH-MS transiting binaries for our assumed
parameters, a formidably small population to find, especially considering that
the signal is at a level of $10^{-3}-10^{-4}$ which may be swamped by stellar
variability.  Space-based transit searches can easily reach this precision over
the transit timescale;  however, they will monitor only a small fraction of the
number of stars required to find BH-MS or NS-MS binaries.  Globular clusters may 
be the best locations to search for these binaries due to the concentration of 
stars and the large over-representation of accreting BH-MS and NS-MS binaries 
relative to the rest of the Galaxy;  however, the precision and time coverage
of current surveys is not sufficient for a detection.

Several microlensing searches are taking place with the goal of finding
deviations in the lightcurve induced by planets \citep{alb95,ben96}.  In the case of a 
terrestrial planet companion of a star lensing a source at the bulge, the Einstein 
radius is $R_E=2.7\times 10^{11} {\rm cm} (M/M_\oplus)^{1/2} (D/4{\rm kpc})^{1/2}$.  
In the case of a Jupiter-mass planet in a globular cluster lensing a source
in the same cluster, the Einstein radius
is $R_E=4.8\times 10^{10} {\rm cm} (M/M_J)^{1/2} (D/1{\rm pc})^{1/2}$ (however,
the optical depth for such events is negligible).
These are somewhat smaller than the size of a red-giant star, which comprise 
a large fraction of microlensing targets given their bright luminosities.  
The region of the microlensing lightcurve near the planet can be treated
as a Chang-Refsdal lens which means that lensing of a red-giant star
by a star-planet system will create fluctuations of the magnitude
given by equation (\ref{cramp}).  This generalizes a result from
\citet{gou97}, who computed the amplification of a large source lensed
by a point mass with shear, to include limb-darkening and convergence.
\citet{gou96} has argued for the monitoring of giant stars to look for
planets given their brightness, lack of confusion, and large cross section.
If such events are found, the limb-darkened lightcurves can be approximated
by equation (\ref{cramp}).

\section{Conclusions}

We have shown that lensing of a large source adds an area which is approximately
constant when the source dimensions are much larger than the Einstein radius, 
which leads to an accurate approximation for the lensing lightcurve in the
small amplification limit, typically $\delta \mu < 0.01$ for this approximation
to be relevant.  We have extended this result to include limb-darkening,
shear, and convergence.  Since many large-area accurate photometric searches are 
being carried out or planned, it is likely that this formalism will be relevant for
planning observations and discovering dark objects via lensing.
The most relevant applications are likely to be lensing by white
dwarfs in binaries and lensing perturbations by planets during Galactic microlensing
events.  Using the formalism of this paper we have rederived a result
that the fluctuations of a distant source lensed by a random star field
at low optical depth varies as the square root of the optical depth times
the ratio of the Einstein radius to the source size, first derived
by \citet{ref91}.

\acknowledgments

Support for E.A. was provided by the National Aeronautics and Space 
Administration
through Chandra Postdoctoral Fellowship Award PF0-10013 issued by the Chandra
X-ray Observatory Center, which is operated by the Smithsonian Astrophysical
Observatory for and on behalf of the National Aeronautics Space Administration
under contract NAS 8-39073.

\end{document}